\documentclass[10pt,twocolumn,letterpaper]{article}

\usepackage{cvpr}
\usepackage{times}
\usepackage{epsfig}
\usepackage{graphicx}
\usepackage{amsmath}
\usepackage{amssymb}

\usepackage{algorithm} 
\usepackage{algorithmic} 
\usepackage{caption}
\usepackage{subfigure}
\usepackage{booktabs}
\usepackage{threeparttable}

\usepackage[pagebackref=true,breaklinks=true,letterpaper=true,colorlinks,bookmarks=false]{hyperref}
\cvprfinalcopy 

\def\cvprPaperID{5} 

\ifcvprfinal\fi
\begin{document}

\title{Learned Video Compression with Feature-level Residuals}

\author{Runsen Feng, Yaojun Wu, Zongyu Guo, Zhizheng Zhang, Xin Jin, Zhibo Chen\thanks{Corresponding author.}\\
{\textit{CAS Key Laboratory of Technology in Geo-spatial Information Processing and Application System}}\\
\textit{University of Science and Technology of China}\\
{\tt\small \{fengruns,yaojunwu,guozy,zhizheng,jinxustc\}@mail.ustc.edu.cn, chenzhibo@ustc.edu.cn}}


\maketitle

\begin{abstract}
In this paper, we present an end-to-end video compression network for P-frame challenge on CLIC. We focus on deep neural network (DNN) based video compression, and improve the current frameworks from three aspects. First, we notice that pixel space residuals is sensitive to the prediction errors of optical flow based motion compensation. To suppress the relative influence, we propose to compress the residuals of image feature rather than the residuals of image pixels. Furthermore, we combine the advantages of both pixel-level and feature-level residual compression methods by model ensembling. Finally, we propose a step-by-step training strategy to improve the training efficiency of the whole framework.
Experiment results indicate that our proposed method achieves 0.9968 MS-SSIM on CLIC validation set and 0.9967 MS-SSIM on test set.

\end{abstract}

\section{Introduction}
Video data has occupied more than 80\% internet transmission resources~\cite{cisco2018cisco} in recent years, and the percentage will be increased even further. As the demand for video transmission grows, traditional video codec has been studied for decades as a way to save internet bandwidth. It typically tackles the compression problem with four basic steps: prediction, transform, quantization and entropy coding. Although it has been developed for many years, it still updates a new generation every 8 to 10 years, and each generation can bring nearly 50\% gain on bandwidth saving.

Recently, deep learning technology opens countless possibilities in further improving the coding performance~\cite{cheng2019learning,wu2018video,lu2019dvc,cheng2019learning,habibian2019video,djelouah2019neural}. Unlike traditional video codec, DNN based methods can jointly optimize their network components in an end-to-end manner, which can further improve the rate-distortion performance of compression~\cite{lu2019dvc}. Among DNN based methods, Chen \etal~\cite{chen2019learning} first propose to predict block of pixels autoregressively, and the residuals is encoded by an autoencoder. Besides, Wu \etal \cite{wu2018video} propose a interpolation-based approach with traditional motion vectors. They recurrently compress the residuals conditioned on warped reference frames and context. Following conventional video coding architecture, Lu \etal~\cite{lu2019dvc} propose an end-to-end trainable framework by replacing all key components in the classical video codec with DNNs. In their framework, optical flow is used for motion compensation, and two autoencoders are adopted to encode the corresponding optical flow and residuals, respectively. 
\cite{djelouah2019neural} perform temporal interpolation by the decoded optical flow and blending coefficients, and directly quantize the latent space residuals by reusing the same autoencoder of image compression. 
Besides, Habibian \etal~\cite{habibian2019video} propose a rate-distortion autoencoder which consists of a 3D autoencoder transform and an autoregressive model for entropy coding in latent space.

In this paper, we propose an end-to-end video compression framework based on \cite{lu2019dvc} for CLIC 2020 P-frame task. We focus on improving the performance of residual coding, which is not fully explored in previous methods. Our contributions can be summarized as follows: 
\begin{itemize}
\item We compress the residual information from the perspective of feature level instead of pixel level, which can effectively suppress the influence of motion compensation error.
\item We involve both of our pixel-level and feature-level residual compression methods within an ensemble model, which further improves the compression performance.
\item To reduce the difficulty of training multiple modules from scratch, we thus design a step-by-step training strategy to improve the training efficiency.
\end{itemize}  

\begin{figure}[t]
\setlength{\abovecaptionskip}{3pt} 
\setlength{\belowcaptionskip}{-1pt}
  \centering
  \includegraphics[width=0.4\textwidth]{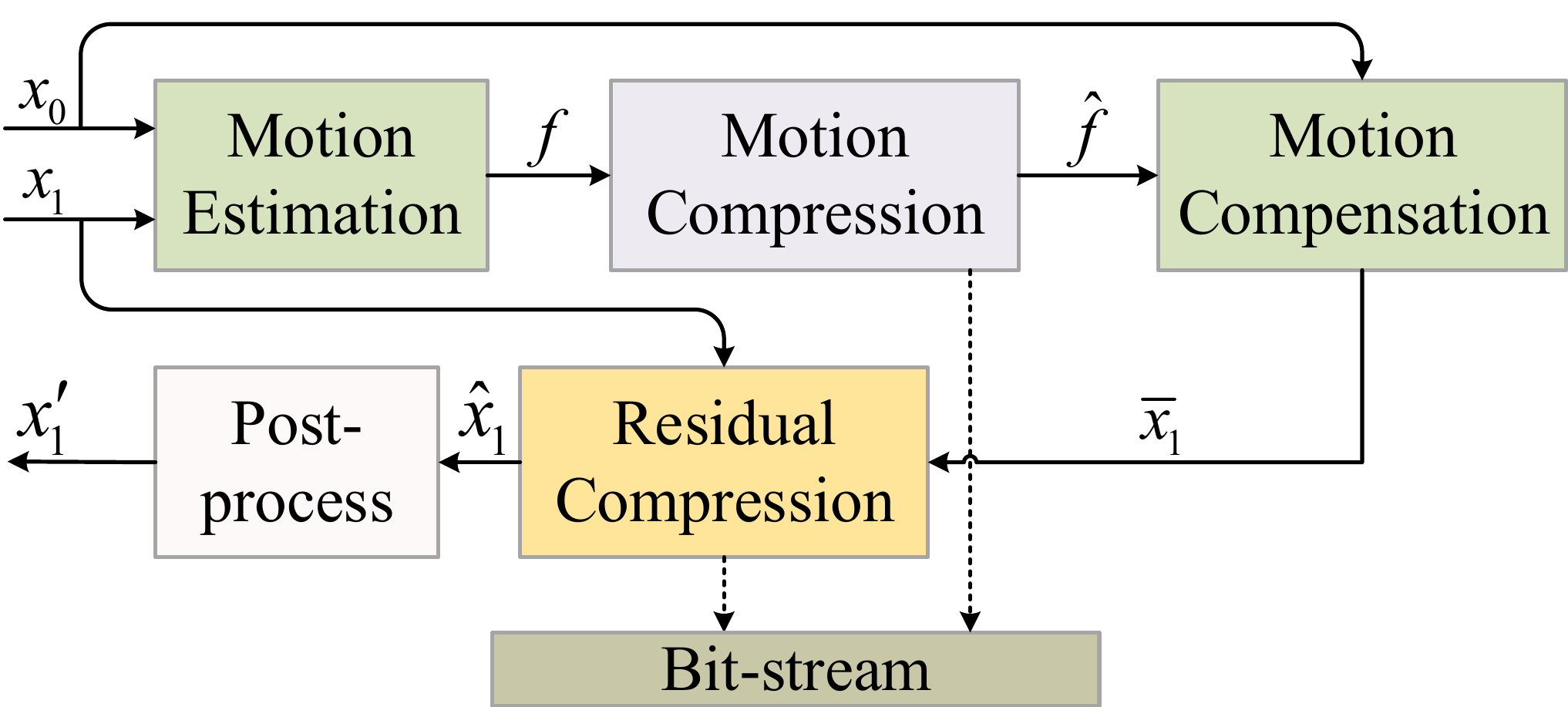}
  \caption{The overall framework of our proposed method.}
  \label{fig:overall}
\end{figure}

\section{Method}
\subsection{Overall Framework}
In this section, we give a high-level overview of our framework. Figure~\ref{fig:overall} shows the overall coding pipeline as well as basic components. Brief summarization on these components is introduced as follows:

\paragraph{Motion Estimation.} We apply PWC-Net~\cite{sun2018pwc} as our motion estimation network to compute optical flow between two adjacent frames. Specifically, the target frame $\boldsymbol{x}_1$ and the reference frame $\boldsymbol{x}_0$ are fed into the motion estimation network to extract motion information $\boldsymbol{f}$, which will be encoded by motion compression module.

\paragraph{Motion Compression.} 
To compress optical flow for transmission, we propose a VAE based motion compression network based on the factorized-prior model proposed by ~\cite{balle2018variational}. We replace GDN with residual blocks and further add a downsampling layer. This is because the spatial redundancy in optical flow is much higher than that in image. Through motion compression, we can obtain reconstructed optical flow $\boldsymbol{\hat{f}}$ for motion compensation.

\paragraph{Motion Compensation.} In motion compensation, we try to obtain prediction frame $\boldsymbol{\overline{x}}_1$ through the decoded optical flow $\boldsymbol{\hat{f}}$ and the reference frame $\boldsymbol{x}_0$. Detailed structure of the motion compensation network can be seen in \cite{lu2019dvc}.

\paragraph{Residual Compression.}
We aim to encode the residual information between the compensation frame $\boldsymbol{\overline{x}}_{1}$ and the original frame $\boldsymbol{x}_1$.  There are two variants of our residual compression networks as shown in \ref{fig:tworesidual}. For the first variant, we directly compute the pixel space residuals which then are compressed as image data. We employ our pixel-level residual compression network based on the hyperprior model in \cite{balle2018variational}. For the second variant, we calculate and compress residuals in feature domain. The corresponding feature-level residual compression network will be described in section~\ref{sec:feature_residual}. 

\paragraph{Post-process}
Inspired by the success of image quality enhancement networks, we add GRDN ~\cite{kim2019grdn} as our post-process network to enhance the quality of reconstructed frame $\boldsymbol{\hat{x}}_1$ after residual compression. The final output of the post-process network is denoted as $\boldsymbol{x}'_{1}$.

\subsection{Feature-level Residuals}
\label{sec:feature_residual} 
\begin{figure}[t]
\setlength{\abovecaptionskip}{0pt} 
\setlength{\belowcaptionskip}{-1pt}
\centering
\subfigure[Residual compression on pixel level.]{
\label{fig:pixel}
\includegraphics[width=0.55\linewidth]{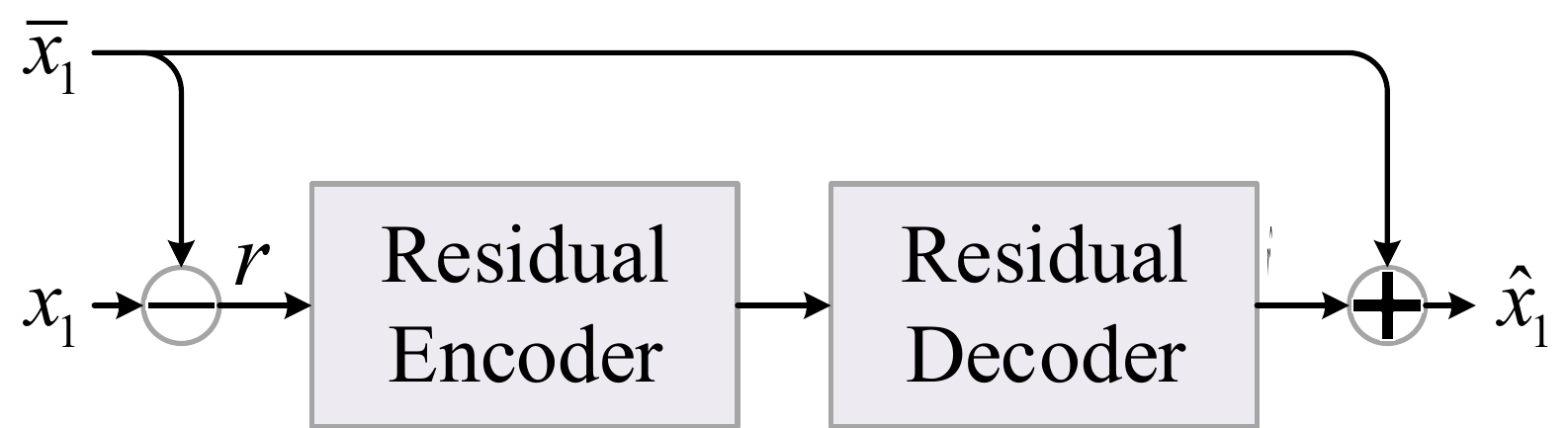}}
  
\subfigure[Residual compression on feature level.]{
\label{fig:flow}
\includegraphics[width=0.90\linewidth]{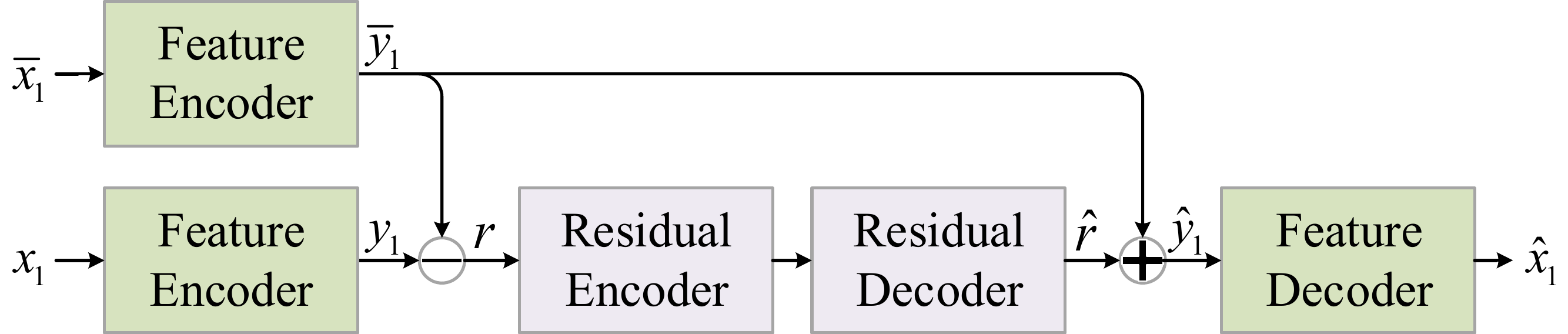}}
\caption{Two residual compression methods in our framework.}
\label{fig:tworesidual}
\end{figure}

Residual compression module plays an important role in exploring the remaining spatial redundancy conditioned on motion compensation frame. Traditional approaches typically calculate and compress the residuals in pixel space. However, the corresponding residuals are easily affected by the prediction error from optical flow based motion compensation. Instead of directly computing pixel-level residuals, we propose to compress the residuals of image feature. As shown in Figure~\ref{fig:flow}, the compensation frame $\boldsymbol{\overline{x}}_1$ and the target frame $\boldsymbol{x}_1$ are transformed into feature space by a shared feature encoder, and then the residual of their feature is compressed by a variational autoencoder. Note that different from the work~\cite{djelouah2019neural} which directly compute and quantize latent space residuals to reuse image compression network, our method calculate residuals in high-dimensional feature space and further transform the residuals into a discrete latent representation. 

\begin{figure*}[t]
\setlength{\abovecaptionskip}{5pt} 
\setlength{\belowcaptionskip}{-1pt}
  \centering
  \includegraphics[width=0.875\textwidth]{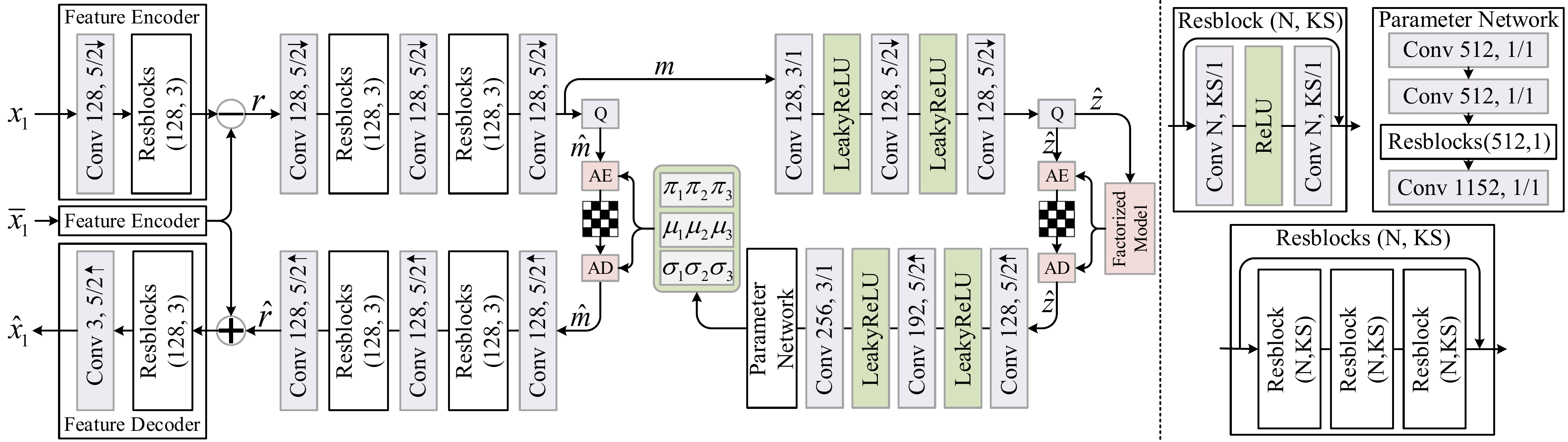}
  \caption{Framework of residual compression in feature space. Q represents quantization and AE, AD respectively represent arithmetic encoder and arithmetic decoder. The parameters of convolution are denoted as numbers of filters, kernel size/down- or upsampling stride, where $\downarrow$ indicates down sampling and $\uparrow$ indicates up sampling.}
  \label{fig:residual_cmp}
\end{figure*}

The detailed structure of the proposed residual compression networks is illustrated in Figure~\ref{fig:residual_cmp}. Similar to exiting work~\cite{balle2018variational} for learned image compression, the VAE for residual compression consists of an autoencoder transform and a hierarchical entropy model. The autoencoder first transforms the feature-level residual $\boldsymbol{r}$ into a quantized latent representation $\boldsymbol{\hat{m}}$ and then maps it back to reconstruction $\boldsymbol{\hat{r}}$. A non-parametric, fully factorized density model proposed by Ball\'{e} \etal~\cite{balle2018variational} is used for estimating the distribution of the hyperprior $\boldsymbol{\hat{z}}$. Conditioned on $\boldsymbol{\hat{z}}$, the distribution of each latent $\hat{m}_i$ is modeled as a Gaussian Mixture convolved with a unit uniform distribution:
\begin{equation}\label{eq:1}
\begin{aligned}
p_{\boldsymbol{\hat{m}}\mid \boldsymbol{\hat{z}}}
(\boldsymbol{\hat{m}}\mid \boldsymbol{\hat{z}})
=
\prod_i{(\sum_{k=1}^{K} \pi_{i, k} \mathcal{N}(\mu_{i, k}, \sigma_{i, k}^2)*\mathcal{U}(-\frac{1}{2}, \frac{1}{2}))}(\hat{m}_i)
\end{aligned}
\end{equation}
where k denotes the index of mixtures. We use $K=3$ in our experiments. The parameters of weight, mean and scale are predicted by the parameter network. Following the work of~\cite{balle2016end}, we replace the rounding function Q (quantization) with additive uniform noise during training.

\subsection{Model Ensemble}
When motion compensation is accurate, feature-level residuals have no more advantage than pixel-level residuals. To further improve the compression performance, we train two individual P-frame frameworks which contain the two residual compression networks shown in Figure~\ref{fig:tworesidual}, respectively. When encoding the validation dataset, both of the frameworks are used to compress each sequence and then an optimal assignment is chosen by a knapsack solver given the target bitrate. Experimental results show that the ensemble of these two frameworks outperforms any of them. 

\subsection{Training Strategy}
\paragraph{Loss Function.} The overall rate-distortion (R-D) loss function can be fomulated as follows: 
\begin{equation}\label{eq:2}
\begin{aligned}
\mathcal{L} = R_f + R_r + \lambda d(\boldsymbol{x}_1, \boldsymbol{x}'_1)
\end{aligned}
\end{equation}
where $R_f$ and $R_r$ represent the rate of optical flow and residual, respectively. The distortion $d$ is measured using MS-SSIM and the coefficient $\lambda$ controls the R-D trade-off.

\paragraph{Step-by-step Training.} 
It is hard to train the whole framework from scratch. Therefore, we divide the training procedure into pre-training stage and fine-tuning stage. In the pre-training stage, we first train the motion compression and motion compensation networks, keeping the parameters of the pre-trained PWC-Net in~\cite{sun2018pwc} fixed. Then the residual compression networks are added for joint training with a large $\lambda$. This is to avoid information loss in pre-training stage. The post-process network is pre-trained on our learned image codec. In the second stage, we jointly fine-tune all of the modules with different $\lambda$. 

\section{Experiments}
\subsection{Implementation Details}
We train our P-frame compression model with two datasets provided by Vimeo-90K\cite{xue2019video} and CLIC, respectively. The Vimeo-90k dataset consists of 89800 clips in an RGB format and the CLIC dataset contains about 466684 frames of YUV420 format. The YUV420 images in the CLIC dataset are converted into YUV444. Each pair of frames are randomly cropped into 256x256 patches during training. The mini-batch size is set as 8. 

In the pre-training stage, we train our model on the Vimeo-90k dataset using the learing rate of $5 \times 10^{-5}$. In the fine-tuning stage, we set learning rate to $1 \times 10^{-5}$ and jointly fine-tune the whole framework on the CLIC dataset. We convert $\boldsymbol{x}_1$ and $\boldsymbol{x}'_1$ back into YUV420 format to calculate distortion when fine-tuning. It takes about 7 days to complete the training procedure using two GTX 1080 Ti GPUs.

The total size in P-frame task, which consists of the compressed data size and model size, should be no more than a target size. Thus, we train our framework with different $\lambda$ and select two most promising models to perform bit-allocation by a knapsack solver.

\subsection{Experimental Results}

\begin{table*}[t]
\setlength{\abovecaptionskip}{0pt} 
\setlength{\belowcaptionskip}{-0pt}
\centering
\caption{Evaluation results on P-frame validation dataset.}
\begin{threeparttable}
\begin{tabular}{cccccccc}
\toprule[1pt]
\begin{tabular}[c]{@{}c@{}}Model\\ \#\end{tabular} & \begin{tabular}[c]{@{}c@{}}Pixel-level \\ residuals\end{tabular} & \begin{tabular}[c]{@{}c@{}}Feature-level\\ residuals\end{tabular} & GRDN & Ensemble & Data size & Model size & MS-SSIM  \\ \hline
\# 1& \checkmark &  &   & & 38205911 & 79114205   & 0.996302 \\ \hline
\# 2& \checkmark &  &\checkmark  & & 37788576 & 120847836  & 0.996619 \\ \hline
\# 3& &\checkmark   &   & & 38133059 & 86399558   & 0.996645 \\ \hline
\# 4& &\checkmark   &\checkmark  & & 37716003 & 128105152  & 0.996700 \\ \hline
\# 5&\checkmark &\checkmark   &\checkmark   & \# 1, \# 4 & 37960950 & 103610323  & 0.996792 \\ \hline
\# 6&\checkmark &\checkmark   &\checkmark   & \# 2, \# 4& 37735411 & 126164272  & 0.996866 \\ \bottomrule[1pt]
\end{tabular}
\begin{tablenotes}
\footnotesize
\item[*] The unit of data/model size is Byte. For P-frame challenge in CLIC, we limit the total size to 3,900,000,000 bytes, which is calculated as follows: model size + 100 $\times$ data size.
\end{tablenotes}
\end{threeparttable}
\label{tab:result}
\end{table*}

We evaluate several design choices on the CLIC validation dataset. Specifically, we evaluate the benefits of the post-process module, the proposed feature-level residuals and the multi-model ensemble. For a fair comparison, two models trained by different $\lambda$ are used to make bit-allocation for each framework . 

Evaluation results are shown in Table~\ref{tab:result}. It is obvious that the proposed feature-level residuals consistently bring performance gain, with or without post-process networks. When GRDN is included, significant performance improvement occurs on the framework of pixel-level residuals. Compared with GRDN, the results in second and third rows indicate that our feature-level residual compression module performs better with less model parameters. The ensemble modeling further improves the performance as shown in the fifth and sixth rows of the table. Due to time reason, we only submit the version which achieves 0.996792 MS-SSIM on validation dataset. Our final ensemble model, which reaches 0.996866 MS-SSIM, is comprised of the models in the second and fourth rows.

To satisfied the limitation of decoding time, the autoregressive entropy model in~\cite{minnen2018joint} is not implemented in our framework. The introduction of autoregressive models can further boost the performance for both motion compression and residual compression modules.

\section{Conclusion}
\setlength{\abovecaptionskip}{0pt} 
\setlength{\belowcaptionskip}{-0pt}
In this paper, we propose an end-to-end video compression framework for the P-frame challenge in CLIC 2020. Firstly, we propose a feature-level residual compression network to reducing the effects of motion compensation error. Secondly, we combine two different residual compression frameworks in an ensemble model to further improve the performance. Lastly, a step-by-step training strategy is proposed to further improve the training efficiency.
As shown in the leaderboard of P-frame validation, our team 'IMCL\_MSSSIM' achieved 0.9968 MS-SSIM score.
In the future, finding an end-to-end structure to jointly exploit both pixel-level and feature-level residuals would be an interesting direction.

\section*{Acknowledgment}
\setlength{\abovecaptionskip}{0pt} 
\setlength{\belowcaptionskip}{-0pt}
This work was supported in part by NSFC under Grant U1908209, 61632001 and the National Key Research and Development Program of China 2018AAA0101400.

{\small
\bibliographystyle{bib/ieee_fullname}
\bibliography{bib/reference}
}

\end{document}